\documentclass[draft,jgrga]{agutex}
\usepackage{txfonts,graphicx,multirow,color}
\setkeys{Gin}{draft=false}

\authorrunninghead{FOLEY ET AL.}

\titlerunninghead{PULSE PROPERTIES OF FERMI-GBM TGFS}

\authoraddr{P.~N.~Bhat, M.~S.~Briggs, V.~L.~Chaplin, V.~Connaughton, R.~D.~Preece, and S.~Xiong, CSPAR, University of Alabama in Huntsville, 320 Sparkman Dr., Huntsville, AL 35805, USA.}

\authoraddr{J.~R.~Dwyer and E.~Cramer, Department of Physics and Space Sciences, Florida Institute of Technology, Melbourne, FL 32901, USA.}

\authoraddr{G.~J.~Fishman, Jacobs Engineering Inc., Huntsville, Alabama, USA.}

\authoraddr{S. Foley, G.~Fitzpatrick, S.~McBreen, and D.~Tierney, School of Physics, University College Dublin, Belfield, Dublin 4, Ireland. (suzanne.foley2@ucd.ie)}

\authoraddr{J.~Greiner, and A.~von Kienlin, Max-Planck-Institut f\"{u}r extraterrestrische Physik, 85748 Garching, Germany.}

\authoraddr{R.~M.~Kippen, ISR-1, Los Alamos National Laboratory, Los Alamos, NM 87545, USA.}

\authoraddr{C.~A.~Meegan and W.~S.~Paciesas, Universities Space Research Association, 320 Sparkman Dr., Huntsville, AL 35805, USA.}

\authoraddr{C.~Wilson-Hodge,  Space Science Office, VP62, NASA Marshall Space Flight Center, Huntsville, AL 35812, USA.}

\begin{document}
\title{Pulse Properties of Terrestrial Gamma-ray Flashes detected by the Fermi Gamma-ray Burst Monitor}

\authors{S. Foley, \altaffilmark{1,2}  G.~Fitzpatrick, \altaffilmark{1}  M.~S.~Briggs, \altaffilmark{3,4} V.~Connaughton, \altaffilmark{3,4} D.~Tierney, \altaffilmark{1} S.~McBreen, \altaffilmark{1} J.~R.~Dwyer, \altaffilmark{5} V.~L.~Chaplin, \altaffilmark{3} P.~N.~Bhat, \altaffilmark{3} D.~Byrne, \altaffilmark{1} E.~Cramer, \altaffilmark{5} G.~J.~Fishman, \altaffilmark{6} S.~Xiong, \altaffilmark{3} J.~Greiner, \altaffilmark{2} R.~M.~Kippen, \altaffilmark{7} C.~A.~Meegan, \altaffilmark{8} W.~S.~Paciesas, \altaffilmark{8} R.~D.~Preece, \altaffilmark{3,4} A.~von~Kienlin, \altaffilmark{2} and C.~Wilson-Hodge \altaffilmark{9}}

\altaffiltext{1}{School of Physics, University College Dublin, Belfield, Dublin 4, Ireland.}

\altaffiltext{2}{Max-Planck-Institut f\"{u}r extraterrestrische Physik, Giessenbachstrasse 1, 85748 Garching, Germany.}

\altaffiltext{3}{Center for Space Plasma and Aeronomic Research, University of Alabama in Huntsville, 320 Sparkman Drive, Huntsville, AL 35805, USA.}

\altaffiltext{4}{Department of Physics, University of Alabama in Huntsville, 320 Sparkman Drive, Huntsville, AL 35805, USA.}

\altaffiltext{5}{Department of Physics and Space Sciences, Florida Institute of Technology, Melbourne, Florida, USA.}

\altaffiltext{6}{Jacobs Engineering Inc., Huntsville, Alabama, USA.}

\altaffiltext{7}{Los Alamos National Laboratory, PO Box 1663, Los Alamos, NM 87545, USA.}

\altaffiltext{8}{Universities Space Research Association, 320 Sparkman Drive, Huntsville, AL 35805, USA.}

\altaffiltext{9}{Space Science Office, VP62, NASA Marshall Space Flight Center, Huntsville, AL 35812, USA.}

\begin{abstract}

The Gamma-ray Burst Monitor (GBM) on board the Fermi Gamma-ray Space Telescope has triggered on over 300 Terrestrial Gamma-ray Flashes (TGFs) since its launch in June 2008. With 14 detectors,  GBM collects on average $\sim$\,100 counts per triggered TGF, enabling unprecedented studies of the time profiles of TGFs. Here we present the first rigorous analysis of the temporal properties of a large sample of TGFs (278), including the distributions of the rise and fall time of the individual pulses and their durations. A variety of time profiles are observed with 19\% of TGFs having multiple pulses separated in time and 31 clear cases of partially overlapping pulses. The effect of instrumental dead time and pulse pile-up on the temporal properties are also presented. As the observed gamma-ray pulse structure is representative of the electron flux at the source, TGF pulse parameters are critical to distinguish between relativistic feedback discharge and lightning leader models. We show that at least 67\% of TGFs at satellite altitudes are significantly asymmetric. For the asymmetric pulses, the rise times are almost always shorter than the fall times. Those which are not are consistent with statistical fluctuations. The median rise time for asymmetric pulses is $\sim$\,3 times shorter than for symmetric pulses while their fall times are comparable. The asymmetric shapes observed are consistent with the Relativistic Feedback Discharge (RFD) model when Compton scattering of photons between the source and Fermi is included and instrumental effects are taken into account.

\end{abstract}

\begin{article}

\section{Introduction}

Terrestrial Gamma-ray Flashes (TGFs) were first observed by the Burst and Transient Source Experiment (BATSE) on board the Compton Gamma-Ray Observatory \citep{fishman:1994} as brief, intense flashes of $\gamma$-ray emission coming from the Earth's atmosphere and are linked to electrical activity in atmospheric regions in or above thunderstorms (e.g., \cite{fishman:1994,smith:2010}).
The leading model for TGF production relies on the relativistic runaway electron avalanche (RREA) process \citep{gurevich:1992,roussel:1996,lehtinen:1999}, in which energetic electrons are accelerated in the presence of the strong electric field of a thunderstorm and produce an avalanche of relativistic electrons, which emit $\gamma$-rays via bremsstrahlung radiation. Positrons and photons created by the runaway avalanche may in turn contribute to producing more energetic electrons in a process known as relativistic feedback, which can more easily explain the observed fluxes~\citep{dwyer:2003,dwyer:2007}. For a recent review of TGFs see \citet{dwyer:2012b}. 

Since their discovery, the properties of these events have been studied with BATSE~\citep{fishman:1994}, the Reuven Ramaty High Energy Solar Spectroscopic Imager (RHESSI) \citep[e.g.,][]{grefenstette:2009}, the Astro-rivelatore Gamma a Immagini LEggero (AGILE) \citep[e.g.,][]{marisaldi:2010a} and the Gamma-ray Burst Monitor (GBM) and Large Area Telescope (LAT) on board the \textit{Fermi} Gamma-ray Space Telescope~\citep[e.g.,][]{briggs:2010,grove:2012}. These observations have shown TGFs to be characterized by very short durations and hard spectra. They have generally been found to consist of one or in some cases multiple pulses of $\gamma$-ray emission, with each pulse having a typical duration of less than 1\,ms. BATSE detected 76 TGFs over 9 years which exhibited varied temporal behaviour, including several TGFs with multiple distinct pulses. The TGF durations ranged from sub-millisecond to several milliseconds~\citep{nemiroff:1997,gjesteland:2010}. However the triggering timescale of 64\,ms used for BATSE caused an observational bias towards multi-pulse and longer events~\citep{grefenstette:2008,gjesteland:2010}. The observed BATSE temporal properties were also shown to have been heavily modified by instrumental effects {\citep{grefenstette:2008,gjesteland:2010}. RHESSI amassed a large sample of 820 TGFs between 2002 and 2008 and shows a much larger proportion of short single-pulse events than BATSE, with almost all having durations $<$1\,ms~\citep{grefenstette:2009}. A new analysis of AGILE observations shows a median duration of $\sim$370\,$\mu$s \citep{marisaldi:2013}. The comparison of the results from different missions is complicated by the intrinsic differences in the detectors used and the various methods used to determine the duration.

Despite these TGF observations, a rigorous analysis of the temporal properties of individual TGF pulses has not yet been performed. This is due to the low counting statistics and/or low temporal resolutions of the above instruments which do not allow such an analysis to be performed. Instrumental effects such as dead time and pulse pile-up also affect the observed temporal properties, further complicating such studies. 

The time profile of the gamma ray emission at the source follows the runaway electron flux. \citet{dwyer:2012} suggests that relativistic feedback naturally makes roughly symmetric TGF pulses, regardless of the pulse duration, and asymmetric pulses may be observed due to Compton scattering as the photons propagate to spacecraft altitude. It has been suggested that TGFs may be produced by mechanisms associated with the stepping of lightning leaders \citep{moss:2006,celestin:2011}. In such a case, short TGF durations ($\sim$50\,$\mu$s) may be consistent with the temporal dispersion associated with the Compton scattering of photons produced by an instantaneous TGF source, and longer TGFs may be due to overlapping emissions. Constraining the shapes and durations of TGF pulses is therefore of utmost importance in distinguishing between TGF production models. 

GBM has triggered on hundreds of TGFs since its launch in June 2008. Despite being optimized for $\gamma$-ray burst observations, GBM is particularly well-suited to TGF studies. The large bismuth germanate (BGO) detectors are particularly important for the detection of the higher-energy photons associated with TGFs and enable the accumulation of a large number of counts. On average, the 14 GBM detectors register $\sim$100 counts per triggered TGF. Combined with an absolute timing accuracy of several $\mu$s, this allows the temporal properties of TGFs to be constrained with unprecedented accuracy. \citet{briggs:2010} found that the pulses in a small sample of 12 GBM TGFs (including at least one unrecognised Terrestrial Electron Beam (TEB)) tend to have symmetric or asymmetric shapes and are well-fit with Gaussian and lognormal functions. \citet{fishman:2011} presented a qualitative overview of the time profiles of the first 50 TGFs triggered by GBM but did not perform a quantitative analysis of individual pulse shapes.

In this paper we perform a rigorous quantitative temporal analysis of TGFs by fitting functions to a large sample of individual TGF pulses. We present the rise times, fall times and durations derived from these pulse fits and quantify pulse asymmetry. We also investigate the impact of instrumental effects on the observed temporal properties of TGF pulses. The sample described in this paper comprises 278 of the first 300 TGFs which triggered the GBM instrument, covering the time period from the launch of \textit{Fermi} in June 2008 up to 06 November 2012. In addition to those TGFs detected by GBM via in-orbit triggering, an offline search for TGFs which did not trigger the instrument has been implemented \citep{briggs:2013}. The TGFs in this new sample are typically much less fluent than the triggered sample (see Figure~4d in \cite{briggs:2013}). The counting statistics required to constrain the fit parameters are not present in most of the untriggered events and we therefore limit our analysis to the sample of triggered GRBs. 

Section~\ref{gbm} describes the GBM instrument. In Section~\ref{analysis}, the pulse-fitting procedure is described. The results are presented in Section~\ref{results} and discussed in Section~\ref{discussion}.

\section{Gamma-ray Burst Monitor}
\label{gbm}

The \textit{Fermi} Gamma-ray Space Telescope was launched in June 2008 and has two instruments on board, the Large Area Telescope (LAT) dedicated to high-energy $\gamma$-ray observations from 20\,MeV to $>$\,300\,GeV~\citep{atwood:2009} and the Gamma-ray Burst Monitor (GBM). GBM is comprised of 12 sodium iodide (NaI) scintillation detectors and two BGO scintillation detectors, covering the energy ranges $\sim$\,8\,keV to 1\,MeV and $\sim$\,200\,keV to 40\,MeV respectively~\citep{meegan:2009}.  The \textit{Fermi} satellite has a nearly circular orbit of 565\,km altitude and an inclination of 25.6\,$^{\circ}$ which means that for a large fraction of its orbit it is located above the tropics, where thunderstorms are prevalent.  

GBM has an on-board triggering process, based on a count rate increase which is statistically significant in two or more detectors within a given timescale and over a particular energy range. In the event of a trigger, high-resolution time-tagged event (TTE) data are telemetered to the ground covering the time period from $\sim$\,30\,s before the trigger time to $\sim$\,300\,s after the trigger time. These data consist of individual photon events in each detector, tagged with their arrival times at 2\,$\mu$s resolution (with an absolute accuracy of several $\mu$s) and associated energy. The rate of TGF triggers is currently $\sim$\,1 every 3\,--\,4 days.  Due to their short durations and extremely high fluxes, the high photon rates of TGFs detected by GBM result in significant dead time losses~\citep{briggs:2010}. The degree to which dead time affects our results is discussed in Section~\ref{analysis_dt}.

\section{Pulse-fitting Analysis}
\label{analysis}

Previous studies of a small sample indicate that GBM TGF pulses tend to be either symmetric or have faster rise times than fall times and are well described by simple Gaussian or lognormal functions~\citep{briggs:2010}. To investigate the true pulse properties of a large TGF sample, a function with greater flexibility is fit to the individual pulses of 278 GBM TGFs, in addition to a constant background rate. The Norris function~\citep{norris:1996} is chosen as it allows for a wide variety of pulse shapes, including those with longer rise times than fall times. It is formulated as:
\begin{equation}
f(t)=\left \{\begin{array}{lr}
R_{\rm max} \exp\left[-\left(\frac{|t-t_p|}{\sigma_r}\right)^{\nu}\right]\textrm{,} & t<t_p \nonumber \\
R_{\rm max} \exp\left[-\left(\frac{|t-t_p|}{\sigma_f}\right)^{\nu}\right]\textrm{,} & t>t_p
\end{array} \right.
\end{equation}
where R$_{\rm max}$ is the maximum rate in counts per second per detector, $t_p$ is the peak time (s), $\sigma_r$ is the rise time constant for $t < t_p$ (s), $\sigma_f$ is the fall time constant for $t > t_p$ (s) and $\nu$ is a measure of the sharpness of the pulse, where lower values indicate a more peaked pulse. A peakedness value of 2 represents a two-sided Gaussian function (the $\sigma$ parameter in the Norris function is related to the $\sigma$ of a Gaussian by a factor of two in the exponent). As it was found that the full set of parameters was typically poorly constrained for weaker TGFs in the sample,  $\nu$ is frozen to 2 for all Norris fits. The best-fit parameters are those which maximise the Poisson likelihood function. A complete description of the analysis process is given in \cite{briggs:2010}.

The rise ($t_r$) and fall ($t_f$) times of the pulses are derived from the fit parameters. For each pulse, $t_r$ and $t_f$ are defined as the time intervals between the 10\% and 90\% flux levels of the rising and falling edges of that pulse respectively, and are given by Eqns.~\ref{eq:norris_tr} and \ref{eq:norris_tf}:
% \begin{linenomath*}
\begin{equation}
t_r=\sigma_r\left[\left(\log\left(\frac{1}{0.1}\right)^{\frac{1}{v}}\right)-\left(\log\left(\frac{1}{0.9}\right)^{\frac{1}{v}}\right)\right]
\label{eq:norris_tr}
\end{equation}
% \end{linenomath*}
% \begin{linenomath*}
\begin{equation}
t_f=\sigma_f\left[\left(\log\left(\frac{1}{0.1}\right)^{\frac{1}{v}}\right)-\left(\log\left(\frac{1}{0.9}\right)^{\frac{1}{v}}\right)\right]
\label{eq:norris_tf}
\end{equation}
% \end{linenomath*}
For $\nu=2$, the rise/fall times are related to the respective $\sigma$ by a multiplicative factor of 1.21. Errors on these pulse properties are obtained by propagating the uncertainties on the function parameters. 
To ensure robust statistics, it is necessary to perform the pulse-fitting analysis over the full energy band of the NaI and BGO detectors. As there is known spectral evolution in TGFs, with a general trend for events to soften over time which has been attributed to Compton Scattering (e.g., \cite{grefenstette:2008,ostgaard:2008}), the pulse parameters will vary depending on the energy interval studied. The derived rise time should be generally speaking, independent of the energy band used, as the rise time interval is dominated by unscattered photons. For the fall time, the energy band is more important, as a lower energy interval should have a longer fall due to the greater proportion of scattered events. Therefore, we stress that the rise times reported here should be representative of the incident rise time, whereas the fall time will be slightly longer than the photon fall time above 1 MeV and slightly shorter than the fall time for $<$1 MeV.

The TGFs described in this paper are those which triggered the GBM instrument and all events have full TTE data coverage. The pulse fits are performed using the combined TTE data of all 12 NaI and 2 BGO detectors over the full GBM energy range (8\,keV to 40\,MeV) at the intrinsic 2\,$\mu$s resolution of the data without dead time or pulse pile-up correction. The \textit{Fermi} GBM data are public and are available at the \textit{Fermi} Science Support Center\footnote{http://\textit{Fermi}.gsfc.nasa.gov/ssc/}. A web catalog of GBM TGF lightcurves is also available\footnote{http://gammaray.nsstc.nasa.gov/gbm/science/tgf}.

The pulses were also fit with normal (symmetric) and lognormal (asymmetric) functions as defined in \citet{briggs:2010}. While the more flexible Norris function allows for a single function to be fit to all TGF pulses, the normal and lognormal fits are also presented here as they have a more physical basis that can be more easily interpreted than the Norris function. It also allows the present results to be compared with previous studies and the results using simpler functions may be useful as the basis for further work or TGF simulations. The criterion for preferring a normal or lognormal fit to the data is based on the Akaike Information Criterion (AIC, \citet{akaike:1974}), which allows for the comparison of the goodness of fit of statistical models. It is defined as:
% \begin{linenomath*}
\begin{equation}
AIC=2\,k-2\,\log\,\mathcal{L},
\end{equation}
% \end{linenomath*}
where k is the number of free parameters in the model and $\mathcal{L}$ is the maximized value of the likelihood function for that model. The model which minimizes the AIC is considered to be the best fit to the data. Therefore if the maximum likelihood is identical for two models, the AIC selects the one defined with the smaller number of parameters. The AIC is also used to determine if an additional pulse is required by the fit for those TGFs with possible multiple overlapping pulses. 

\subsection{Dead time and Pulse Pile-up}
\label{analysis_dt}

As the data we fit are subject to dead time and pulse pile-up losses \citep{chaplin:2012}, the peaks of the of the TGF pulses presented here are underestimated. Using the BGO detectors, \citet{tierney:submitted} performed corrections for dead time, pulse pile-up and detection efficiency in a model independent manner for 106 TGFs detected by GBM. They concluded that the primary observational effect was on the observed maximum rate and fluence, and that temporal properties were not significantly affected. However, that work did not use the NaI detectors, for which the effect of pile-up is greater due to the higher proportion of overflow counts in the NaI data compared to the BGO data (overflow counts, which are above the maximum digitised energy, incur a dead time of 10\,$\mu s$).

In order to quantify the degree to which the observed parameters of TGFs detected by GBM are influenced by instrumental effects, we have performed Monte Carlo simulations. Eight representative TGF pulse shapes (4 symmetric and 4 asymmetric) of varying duration and intensity were each simulated 1000 times. The resulting synthetic TGFs were then passed through a filter which emulates the detection process and includes the effects of dead time and pulse pile-up. The pulses were then fit with the Norris function and the observed pulse parameters compared to the input.  A full description of the simulation process and analysis is given in Appendix\,\ref{instrumental}.

As expected, the observed peak rate was heavily modified for both symmetric and asymmetric TGFs, with the observed value ranging from $\sim$\,10 -- 50\,\% of the input depending on both the duration and initial rate. For the temporal parameters, the mean values of the measured distributions are broadly consistent with the input values, with a tendency to shift to longer times. This shift is generally small, with the input value usually contained within $\pm$1\,$\sigma$ of the mean of the dead time filtered distribution. The exception is the fall times for the asymmetric TGFs, which show the same trend but to a greater degree. This elongation has the effect of artificially increasing the observed duration. However, the effect is small, even for very asymmetric high-rate TGFs, with the mean of the observed duration distribution typically differing from the input value by 20--30\,$\mu s$. These results are in agreement with those of \citet{tierney:submitted}.

\section{Results}
\label{results}

The  $T_{90}$ duration measure has previously been used for TGF pulses \citep[e.g.,][]{briggs:2010}), which is the time it takes to observe 90\% of the counts in a pulse, starting and ending when 5\% and 95\% of the total counts have been observed. This measure is highly dependent on an accurate estimation of the background level and is impossible when there is pulse confusion for overlapping pulses. Additionally, as the $T_{90}$ is modified by dead time, the observed $T_{90}$ should be considered an upper limit on the true duration. Based on the Norris pulse fits presented in this paper, a new TGF pulse duration measure is proposed, defined as the 5\% to 95\% level integration of the pulse fit. Figure\,\ref{fig:durations} shows the $T_{90}$ value as a function of the duration derived from the pulse fit. The proposed duration measure agrees with the $T_{90}$ values measured for isolated pulses for the 226 TGF pulses for which both duration measures were available. An advantage of this new measure is that the effect of dead time can be estimated from simulations. The simulations presented in Appendix\,\ref{instrumental} show that for the highest-rate TGF simulated, the effect of dead time is to increase the duration by $\sim$\,10~\%. 

Of the first 300 TGF-like events detected in orbit by the GBM flight  software, 22 are single-pulsed events with durations derived from the Norris pulse fits of longer than 1\,ms.  Of these, 8 are confirmed TEBs with an additional very likely TEB. Since the underlying physical processes producing TEBs are thought to be understood~\citep{dwyer:2008,briggs:2011}, we exclude these 9 events from the subsequent pulse-fitting analysis. The weaker long events are more difficult to classify; many are likely to be TEBs. In order to avoid contaminating the sample with unidentified TEBs, these 13 events are also excluded from the analysis.

Of the remaining 278 TGFs, 55 have multiple emission episodes, consisting of two or more pulses. The maximum number of statistically significant pulses observed in a single TGF is three, with nine TGFs having three resolvable pulses. There are 31 instances of overlapping pulses which are statistically significant. Of course, we cannot exclude the possibility of other multi-pulse TGFs in the sample for which additional pulses cannot be resolved. An example of a TGF with multiple and overlapping pulses fit with the Norris function is shown in Figure\,\ref{fig:multipulse}.

The pulse-fitting analysis results in 348 individual pulses, to which two cuts were applied. For overlapping pulses the fit parameters are correlated and ill-constrained. A cut removing all of these was applied, leaving 280 isolated pulses. The second removed pulses which had poorly constrained rise and fall times (defined as uncertainties greater than 50\% of the value).  This resulted in a total of 250 isolated pulses with well constrained parameters. 

A scatter plot of rise time against fall time obtained from the Norris fits is plotted in Figure\,\ref{fig:rise_fall_scatter_norris} for all of the pulses that passed the cuts. In this and all subsequent figures, the error bars indicate the 68\% confidence intervals. In general, the rise time is shorter than or comparable to the fall time for TGF pulses. A similar plot is shown in Figure\,\ref{fig:rise_fall_scatter} for pulses fit with Gaussian and lognormal functions. Consequently the sample is divided into those which are consistent with symmetry (33\%; red on Figures \ref{fig:rise_fall_scatter_norris} and \ref{fig:rise_fall_hists}), and those which are asymmetric (67\%). Here we define an event as consistent with symmetry if the interval corresponding to the rise time $\pm$ its uncertainty overlaps with that defined by the fall time $\pm$ its uncertainty. Otherwise, the event is considered to be asymmetric.

Sixty-three percent of the total sample have shorter rise times than fall times (blue), with just 4\% of the sample having longer rise than fall times (green). The distributions of the rise times and fall times derived from the parameters of the pulse fits are shown in Figure\,\ref{fig:rise_fall_hists}.  It is evident that the rise time distributions for the symmetric and asymmetric pulses with $t_r<t_f$ are considerably different, with median values of 116\,$\mu$s and 43\,$\mu$s, respectively. The fall times are comparable for symmetric and asymmetric pulses, having median values of 122\,$\mu$s and 120\,$\mu$s, respectively. 

Just ten pulses have longer rise times than fall times (green on Figures \ref{fig:rise_fall_scatter_norris} and \ref{fig:rise_fall_hists}) . Most of these are consistent with statistical fluctuations from the symmetric group as discussed in Appendix \ref{appendix_asym}. One or two may be due to unrecognized overlapping pulses that were fit with a single Norris pulse. 

The temporal variability of TGF pulses derived from the pulse parameters can also be used to constrain the source emission radius~\citep{dwyer:2008b}. A source with characteristic dimension \textit{R} cannot vary faster than $t\sim R/c$ due to the time required for a `turn-on' signal to propagate and turn on 100\% of the source \citep{carroll:2006}. Various models predict different propagation velocities at the source and therefore different source radii. Using the rise time as the shortest temporal variation of a pulse, an upper limit can be placed on the source radius. Table~\ref{table:source_regions} gives the derived upper limits on the source radii for the three pulses with the shortest rise times for the speed of light and for the typical propagation velocity of a lightning stepped leader. For a lightning stepped leader, there are two velocities to consider, the speed at which the lightning extends while it is advancing and the average velocity of the extension of the lightning which includes the many pauses and steps. In this analysis, we are interested in the former, which we take as $\sim1\times10^7$ m/s \citep[p. 135]{rakov:2003}. The fastest rise time of 7\,$\mu$s was measured for TGF\,101230.452 and gives an upper limit for the source radius of $\sim$2\,km, assuming the speed of light \citep{briggs:2010} and 70\,m assuming the propagation velocity of a lightning stepped leader. To explore the possibility that a TGF with a longer rise time could be modified by dead time or otherwise by the measurement process appear to have a 7\,$\mu$s rise time, we performed 1000 simulations of a TGF with $\sigma_r$ set to 15\,$\mu$s (see \S~\ref{params}). Analysis of these simulations show that the probability of observing a 7\,$\mu$s rise time from such a TGF is $<$0.5\%. Therefore, we conclude that it appears unlikely that TGF~101230.452 has an incident rise time which is substantially longer than the observed 7\,$\mu$s.

\section{Discussion}
\label{discussion}

Assuming that most TGFs are generated inside thunderclouds, there are two classes of TGF models that we consider: 1) lightning leader emissions, either in association with large scale thundercloud electric fields or fields generated by the lightning leader, and 2) the relativistic feedback discharge (RFD) model.  The former involves energetic electron emission from the high fields associated with the lightning leader.  Based upon this mechanism, \citet{celestin:2012} suggest that TGFs are the result of either single or multiple overlapping pulses, with durations of about 10 $\mu$s, that are smeared out by Compton scattering in the atmosphere.  On the other hand, for the RFD model \citep{dwyer:2012}, the production of relativistic runaway electron avalanches becomes self-sustaining through the generation of backward propagating runaway positrons and back-scattered x-rays. 

The TGFs detected by GBM exhibit varied temporal behavior and 19\% of the sample consist of multiple pulses, which may be overlapping or isolated by up to several ms in some cases.  The overlapping pulses seen in 31 TGFs in the sample are evidence that the TGF production process is capable of producing multiple pulses on very short timescales. The shortest resolvable peak separation time observed for overlapping GBM pulses is $\sim200\,\mu$\,s.  Multiple TGF pulses may be signatures of multiple lightning discharges. Simulations of the RFD model show that multiple pulses are routinely produced but requires a large potential difference in the avalanche region~\citep{dwyer:2012}. Overlapping pulses caused by longer feedback times in a pulse train are also predicted by the model. Due to the larger potential required for multiple events, they are predicted to be rare. However, this GBM sample is comprised of those TGFs which triggered the GBM detector over a timescale of 16\,ms and so is biased towards multiple events.

Using simulations we have shown that the temporal parameters are relatively unaffected by dead time and pulse pile-up  (Appendix\,\ref{params}). The duration derived from the pulse parameters should be considered an upper limit, with the true duration being $\sim10$\% shorter. With these caveats in place, the temporal parameters presented here can be used to test theoretical models. For a model in which the applicable velocity is the advancement of a lighting leader, the source radius in constrained to be $\lesssim$ 100 m (Table~\ref{table:source_regions}). As this limit is comparable to the size of a stepped leader, it is easily compatible with lightning leader models for TGFs. For the model in which TGFs are the result of lightning leaders propagating into a high field region and initiating RREA multiplication [Dwyer, 2008], the size limit suggests that only one lightning channel is involved in the gamma-ray emission.  However, as TGFs propagate at the speed of light, this constraint is weakened if several leaders emit simultaneously (closer in time than the minimum observed variability time-scale). Whether or not multiple channels would be distinguishable depends on the relative geometry of the branches and the location of \textit{Fermi}. For the fastest rise times observed by GBM, multiple channels would be indistinguishable if the paths from the sources to GBM differ in length by $\lesssim 2$\,km.

\citet{dwyer:2012} modelled the relativistic feedback processes self-consistently and found that,  for the cases investigated,  longer TGFs often make fewer runaway electrons than shorter pulses. The rise time at spacecraft altitude is less affected by Compton scattering than the duration and is relatively unaffected by dead time (Appendix\,\ref{params}). Therefore in Figure\,\ref{fig:fluence_plots} the rise time is plotted as a function of the observed pulse fluence. There is a slight tendency for pulses with longer rise times to have lower fluences, in qualitative agreement with the RFD model. However, pulses with shorter rise times have fluences which span the observed range. Drawing conclusions from this is difficult, as the observed fluence is modified both by dead time losses and the unknown GBM-source geometry.  The range of observed durations of TGFs has been attributed to short overlapping emissions at the source, which are then smeared out by Compton scattering \citep{celestin:2012}. Any such model would have to explain the symmetric pulses observed by GBM and the significantly shorter rise times observed for asymmetric pulses.

The RFD model predicts that the time profile is always roughly symmetric at the source and at higher energies ($>$\,1\,MeV) at spacecraft altitude. At lower energies the propagation of photons through the atmosphere will cause an elongation in time due to Compton scattering. 
The rise times of the asymmetric pulses are significantly shorter than the symmetric pulses while their fall times are comparable. The observed rise and fall times of GBM TGFs can be explained by the RFD model if, at the source, pulses are intrinsically symmetric and have a range of durations. In this scenario short TGFs have tail emission which is easily observed and are detected as asymmetric pulses with short rise times. The Compton tails of longer pulses may be masked by the main pulse emission, particularly for weak TGFs with fewer statistics, and are observed as symmetric pulses with longer rise times.

The fall times observed range from $\sim$50 -- 200\,$\mu$s. The observed fall time is the combination of the intrinsic fall time of the pulse and an additional Compton scattering component. The tail emission is also observed to be longer due to dead time since the peak emission is suppressed more than the tail, which increases the observed fall time of asymmetric TGFs by $\sim$10\% (Figure\,\ref{dt_fall} Appendix\,\ref{params}). For a given range of altitudes and source\,--\,detector geometries, \citet{ostgaard:2008} showed that time delays on the order of 100\,$\mu$s are expected from Compton scattering. With the inclusion of instrumental effects, the observed asymmetry can therefore be explained by Compton scattering.

\begin{acknowledgments}
We thank the anonymous reviewers for their insightful comments. S.~F. acknowledges the support of the Irish Research Council for Science, Engineering and Technology, cofunded by Marie Curie Actions under FP7. G.~F. acknowledges the support of the Irish Research Council. The \textit{Fermi} GBM Collaboration acknowledges support for GBM development, operations, and data analysis from NASA in the United States and from BMWi/DLR in Germany. M.~S.~B., V.~C. and S.~X. acknowledge support from the \textit{Fermi} Guest Investigator Program. D.~T. acknowledges support from Science Foundation Ireland under grant number 09-RFP-AST-2400. SMB acknowledges support from Science Foundation Ireland under grant number 12/IP/1288. The work by J.~D. has been supported in part by DARPA grant HR0011-1-10-1-0061. All GBM data used in this paper are available at http://fermi.gsfc.nasa.gov/ssc/data/access/gbm/.
\end{acknowledgments}

\appendix
\section{Appendix}\label{instrumental}
We use Monte Carlo simulations to address issues in the data that require further examination. Firstly, the degree to which the observed parameters are affected by dead time must be quantified. We also test if pulses with longer rise times than fall times can be explained by statistical fluctuations. These are addressed in turn below.

\subsection{Simulation method} 
For the purposes of simulation, the TGFs were divided into background and pulse components. The background was simulated as a homogeneous Poisson process, with a rate which was set to the mean of the observed background rate distribution for the NaI and BGO detectors, which is 1.1 and 1.6\,KHz respectively.  Each count was assigned a corresponding energy value which was randomly drawn from the observed background count spectrum distribution. 

The pulse component was simulated by first assuming a particular pulse profile and then using the thinning method for a non-homogeneous Poisson process \citep{lewis:1978}.  Each pulse count was assigned a corresponding energy from a spectrum which was derived by propagating the photons resulting from the RREA of an instantaneous distribution of electrons in an electric field of 400 kV/m from an altitude of 15\,km to 565\,km in a 45$^{\circ}$ beam. This simulated photons were integrated in an annulus of 25\,km at a sub-satellite to source distance of 200\,km, corresponding to the maximum probability of TGF detection (\cite{briggs:2013} and references therein). The NaI energy values were drawn in the range 4 keV - $\infty$ and the BGO energy values were drawn from 100 keV - $\infty$. The pulse and background components were then combined. The combined time profile could then be passed through a filter which emulates the detection process. This filter includes effects from both energy-dependant dead time and pulse pile-up effects.

\subsection{Effect of dead time on observed parameters\label{params}}
Simulations were performed of symmetric and asymmetric time-profile TGFs  with different durations and pulse shapes. For each set of temporal properties, two different peak rates were simulated, one representative of the population and one typical of the brightest events. These are referred to as `representative' and `bright' in the text. In total, 8 pseudo TGFs were simulated.

For the symmetric profiles, the parameters were set to $\sigma_r$=$\sigma_f$=100\,$\mu s$ and $\sigma_r$=$\sigma_f$=200\,$\mu s$ for peak rates of 75 and 100\,KHz. For the asymmetric profiles, the parameters were set to $\sigma_r$=25\,$\mu s$, $\sigma_f$=125\,$\mu s$ and $\sigma_r$=50\,$\mu s$, $\sigma_f$=100\,$\mu s$ for peak rates of 150 and 200\,KHz. Each TGF was simulated and passed through the instrumental filter 1000 times. The resulting TGFs were pulse fit and the distribution of observed parameters compared to the input values. 

The observed peak rate and count fluences were heavily modified for both symmetric and asymmetric TGFs, with the observed value typically differing from the input value by $\sim$\,10 -- 30\,\% depending on both the duration and initial rate. Due to their greater rates, the effect is more pronounced for the asymmetric TGFs. 

For the temporal parameters, the means of the measured distributions are broadly consistent with the input values, with a tendency to shift to longer times. This shift is generally small, with the input value usually contained within $\pm$1\,$\sigma$ of the mean. The exception is the fall times for the asymmetric TGFs, which show the same trend but to a greater degree. This is illustrated in Figures\,\ref{fig:dt_rise} and \ref{fig:dt_fall} which show the observed rise and fall time distributions respectively for a strong asymmetric TGF with the input value represented by a vertical black line.

This elongation has the effect of artificially increasing the observed duration, however, the effect is small, even for very asymmetric high-rate TGFs, with the mean of the observed duration distribution typically differing from the input value by 20 -- 30\,$\mu s$. This is highlighted in Figure\,\ref{fig:dt_duration}, which shows the observed distribution for an bright TGF with an asymmetric time profile. These results are consistent with that of \citet{tierney:submitted}, which showed that the observed temporal properties were not significantly effected by dead time and pulse pile-up.

Another point of interest is the extent to which dead time can change a pulse from symmetric to asymmetric and vice-versa. For the asymmetric TGFs that were strongly asymmetric ($5\times \sigma _r = \sigma_f$) the effect was very small, with less than 1\,\% changing from asymmetric to symmetric. For those which were only slightly asymmetric ($2\times \sigma _r = \sigma_f$) the effect was more significant, with approximately 2\,\% of the representative TGFs and 5\,\% of the bright TGFs changing from asymmetric to symmetric. For the symmetric TGFs, the discussion is slightly more nuanced. Due to statistical fluctuations (discussed in more detail in the following section), only 75\% of the sample before dead time is applied are consistent with symmetry. Once dead time filter is applied and the TGFs refit, this increases to 80\%. 

Several of the observed TGFs exhibit extremely fast temporal variation, with rise times less than 10\,$\mu$s. For the simulations discussed above, the incident and observed rise times were found to be in reasonable agreement. However, it is possible is that a bias exists for shorter timescales and these short events are due to TGFs with longer rise times that have been modified by dead time. To explore this possibility, we consider the TGF with the shortest rise time, 101230.452, which has an observed $\sigma_r$ and $\sigma_f$ of 6 and 71 $\mu$s respectively. Scaling the peak rate to 250\,kHz and setting  $\sigma_r$ to 15\,$\mu$s, we simulate and pulse fit this profile 1000 times. The distribution of rise times are consistent with a Gaussian. Based on the fit parameters, we can conclude that the probability of obtaining a 7 rise time from the distribution is $\sim$0.5\%. The distribution of rise times and the Gaussian fit are shown in Figure~\ref{fig:sim_rise}.

\subsection{Significance of pulses with longer rises than falls}\label{appendix_asym} 
Approximately 5\,\% of the pulse fits to the data result in longer rise times than fall times, that are not consistent with symmetry within the rise and fall time uncertainties. To determine whether this could be explained by statistical fluctuations from the symmetric group, 20 sets of pulse parameters were randomly selected from the sub-sample of fits consistent with symmetry. Each of these TGFs were then simulated 100 times and pulse fit. Figure\,\ref{fig:sim_riseFall} shows the resulting rise time vs fall time.  Approximately 12\,\% of the sample have longer rise time than fall time. The equivalent value for the data is $\sim$\,15\,\%, mostly consistent with the simulations. This implies that the majority of the  TGFs pulse fits with longer rise than fall times are consistent with statistical fluctuations from the symmetric group. The remainder are likely unrecognised overlapping pulses that were fit with a single Norris pulse. 

}

\end{article}
\newpage

\begin{figure}
\includegraphics[width=40pc]{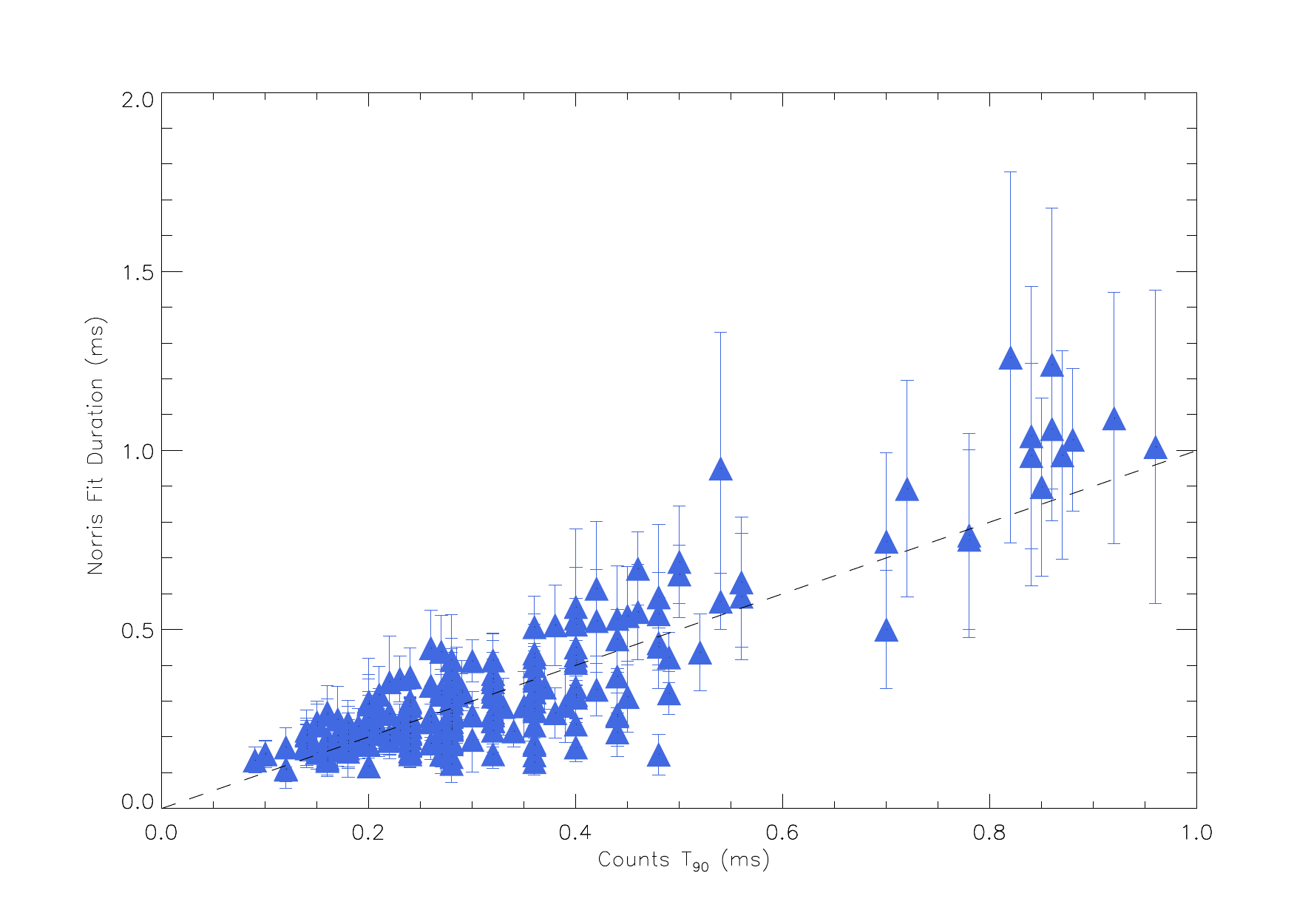}
\caption{T$_{90}$ duration calculated from the count data versus that derived from the pulse fits. The values from the two methods are consistent.}
\label{fig:durations}
\end{figure}

\begin{figure}
\includegraphics[width=40pc]{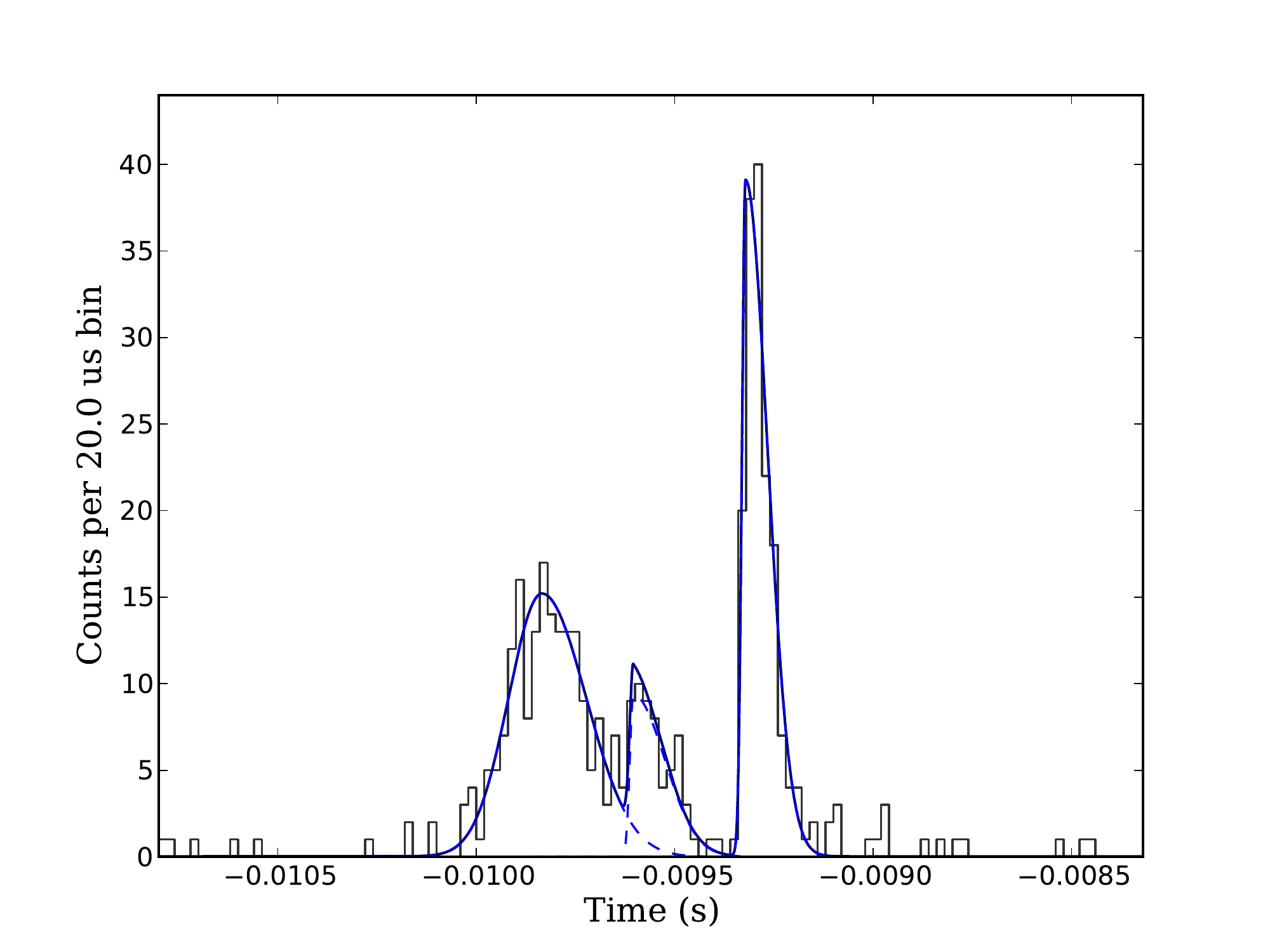}
\caption{GBM TGF110828.435 consisting of multiple and overlapping pulses.  The individual Norris pulse fits are shown as dotted lines; the summed fit is shown as a solid line. Since the fits are performed at the native resolution of 2\,$\mu$s, the choice of data binning for the lightcurve is for illustration purposes only and does not have any effect on the pulse fit.}
\label{fig:multipulse}
\end{figure}	

\begin{figure}
\includegraphics[width=40pc]{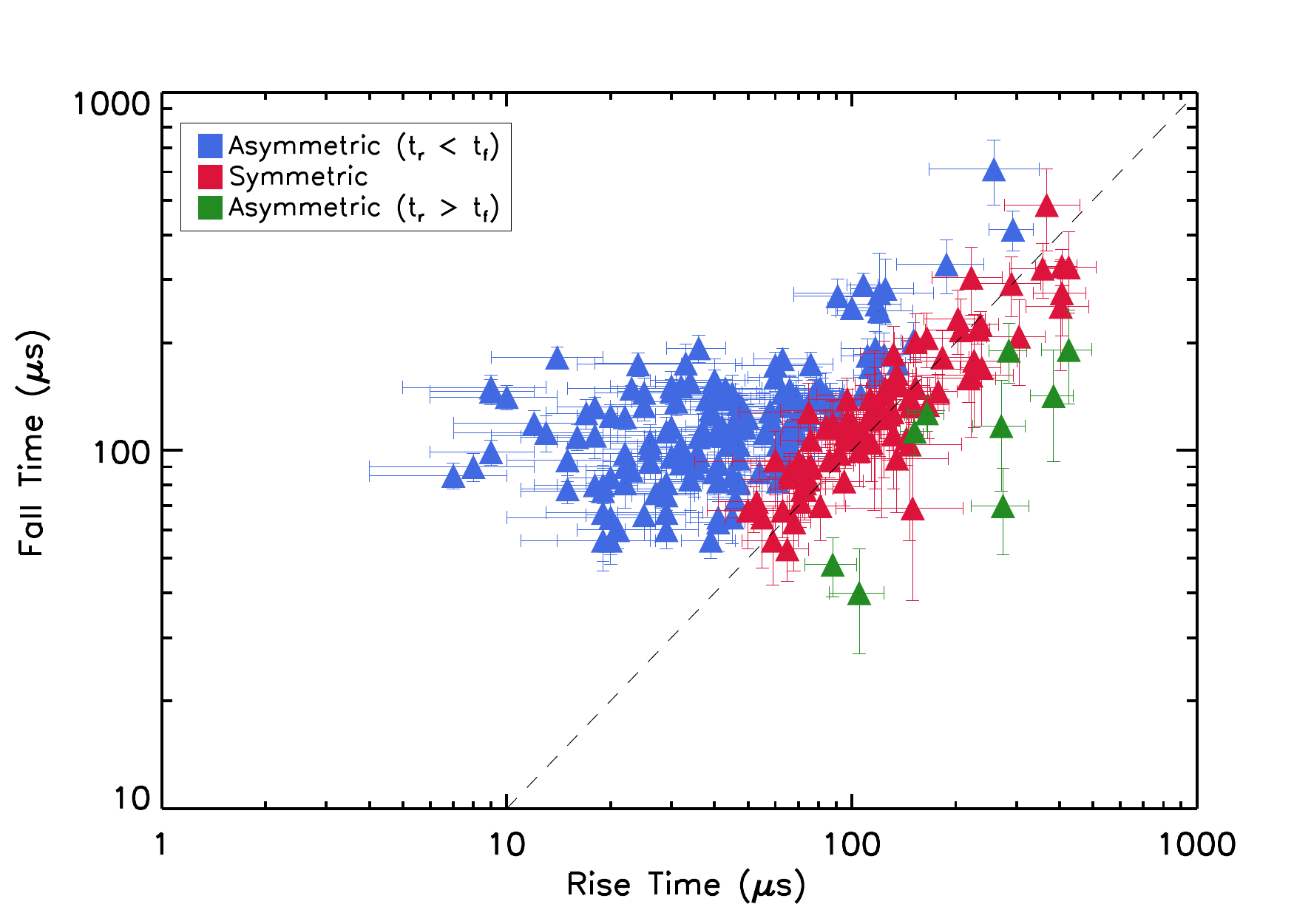}
\caption{Rise time versus fall time for 250 pulses fit with Norris functions. The dashed line represents the line of equal rise times and fall times.}
\label{fig:rise_fall_scatter_norris}
\end{figure}

\begin{figure}
\includegraphics[width=40pc]{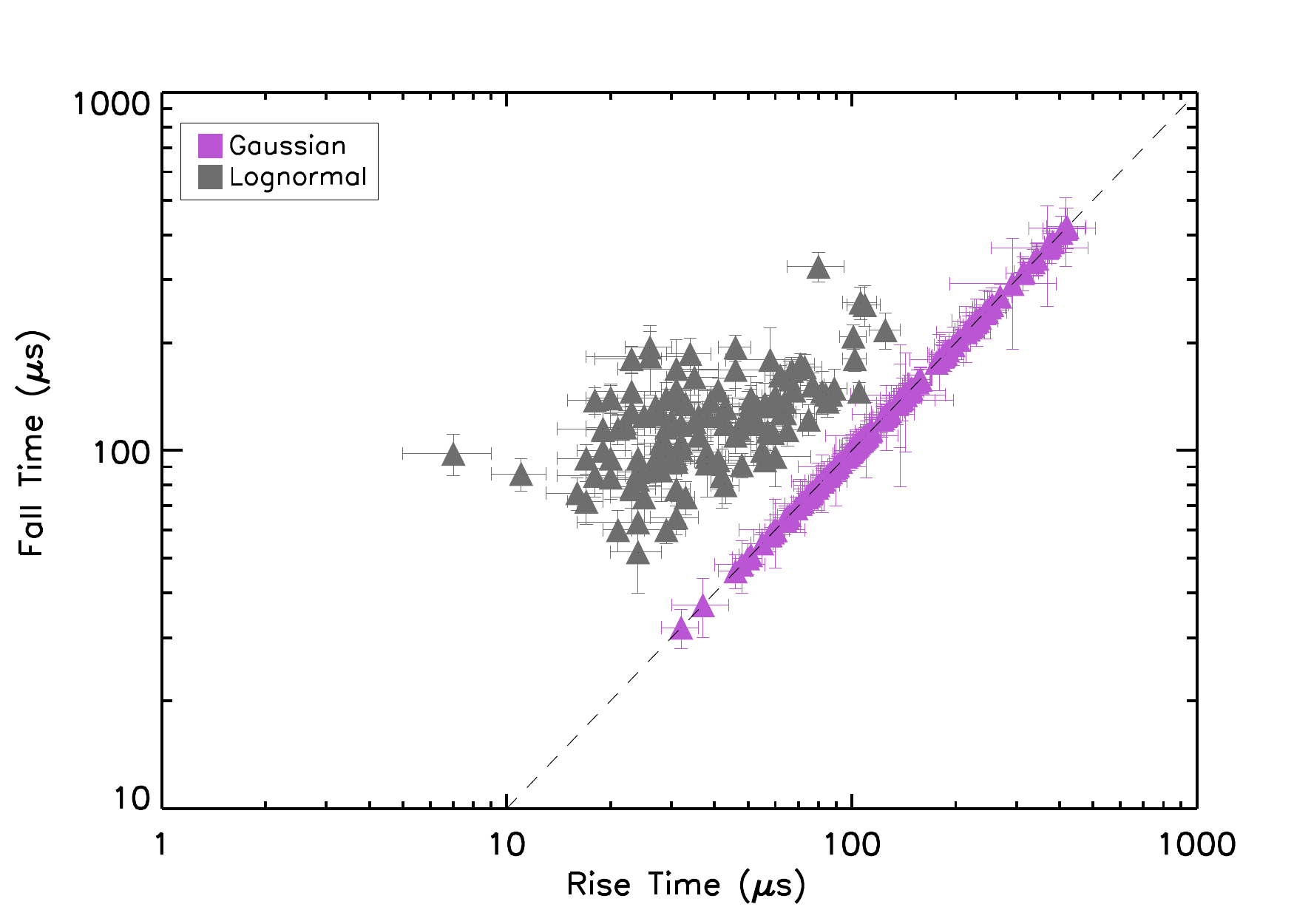}
\caption{Rise time versus fall time for pulses fit with Gaussian functions (purple) and pulses best fit with lognormal functions (gray). The dashed line represents the line of equal rise times and fall times. The pulse parameters obtained from the simpler Gaussian and lognormal fits are better constrained than those derived from the Norris function.}
\label{fig:rise_fall_scatter}
\end{figure}

\begin{figure}
\includegraphics[width=40pc]{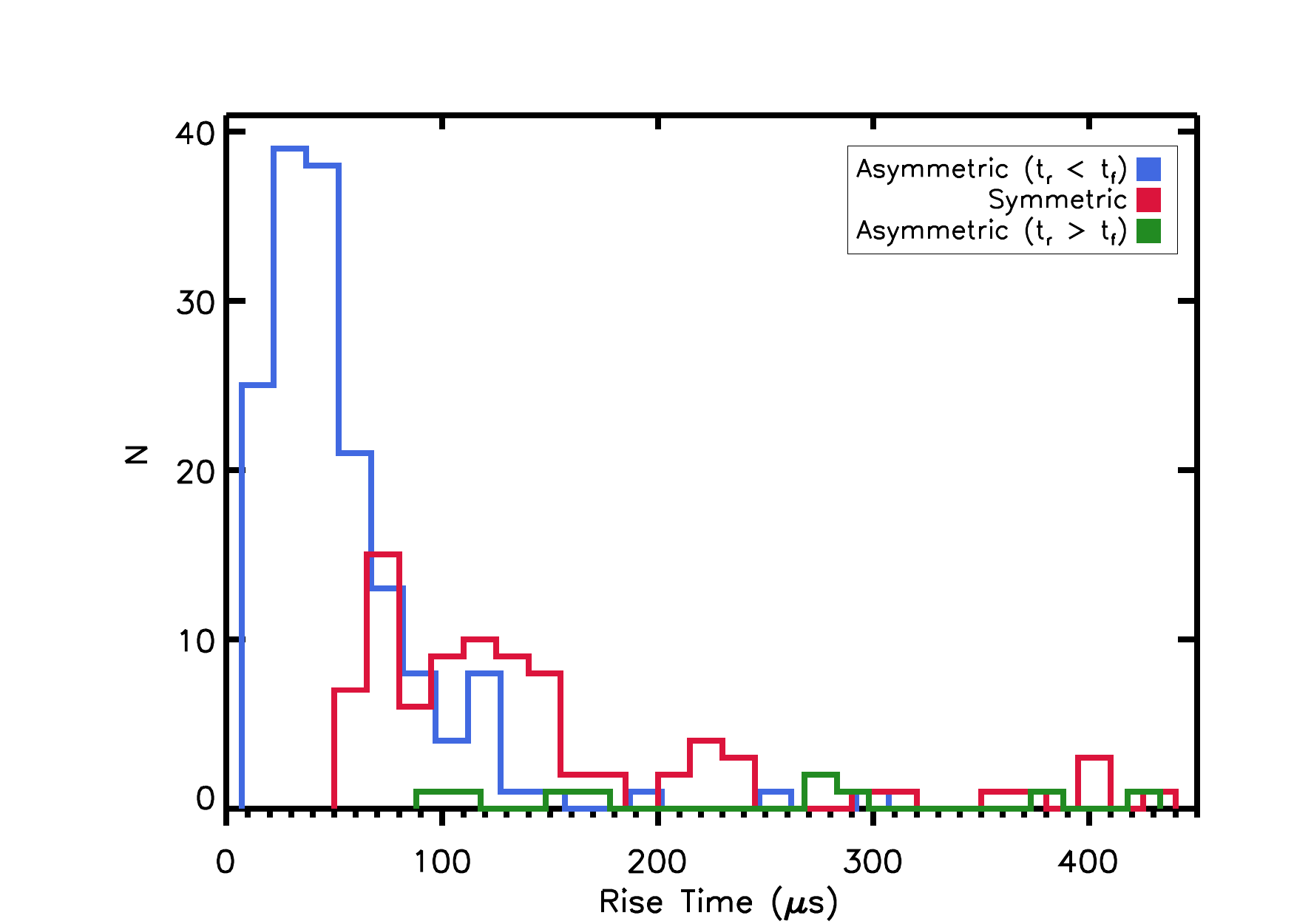}
\includegraphics[width=40pc]{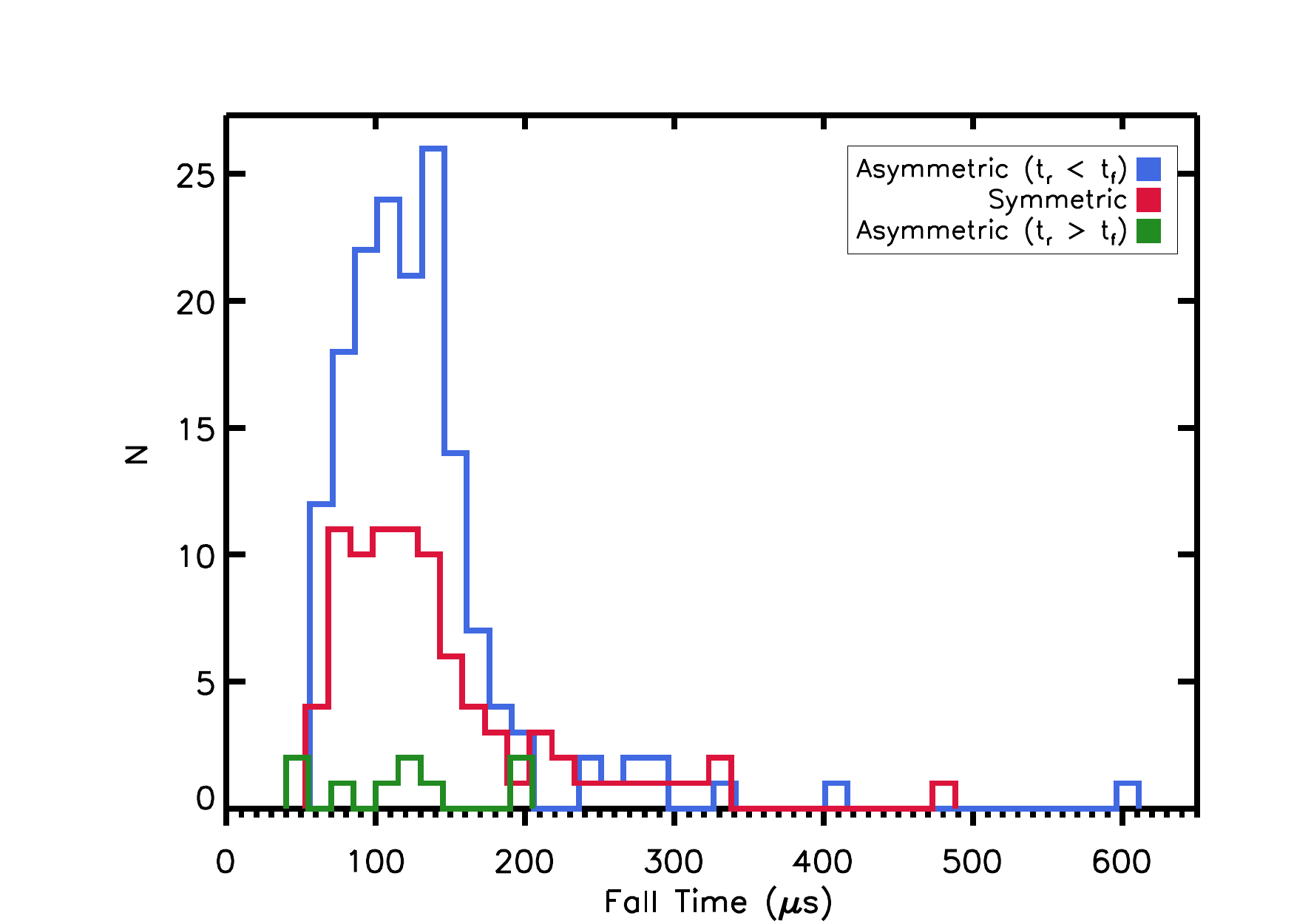}
\caption{Rise time (top) and fall time (bottom) distributions resulting from the Norris fits to a total of 250 TGF pulses.}
\label{fig:rise_fall_hists}
\end{figure}

\begin{table}
\caption{Constraints on the source emission radii for the 3 TGF pulses in the sample with the shortest rise times assuming different propagation velocities. }
\label{table:source_regions}
\begin{center}
\begin{tabular}{@{}l c l c@{}}
\hline
TGF & Rise Time ($\mu$s) & Propagation & Maximum Source Radius (m) \\
\hline
TGF\,101230.452 & $7\pm3$ & Speed of light ($3\times10^8$ m/s) & $2100\pm900$ \\
& & Stepped lightning leaders ($\sim1\times10^7$ m/s) & $70\pm30$ \\
\hline
TGF\,100716.443 & $8\pm4$ & Speed of light ($3\times10^8$ m/s) & $2400\pm1200$ \\
& & Stepped lightning leaders ($\sim1\times10^7$ m/s) & $80\pm40$ \\
\hline
TGF\,100426.657 & $9\pm3$ & Speed of light ($3\times10^8$ m/s) & $2700\pm900$ \\
& & Stepped lightning leaders ($\sim1\times10^7$ m/s) & $90\pm30$ \\
\hline
\end{tabular}
\end{center}
\end{table}

\begin{figure}
\includegraphics[width=40pc]{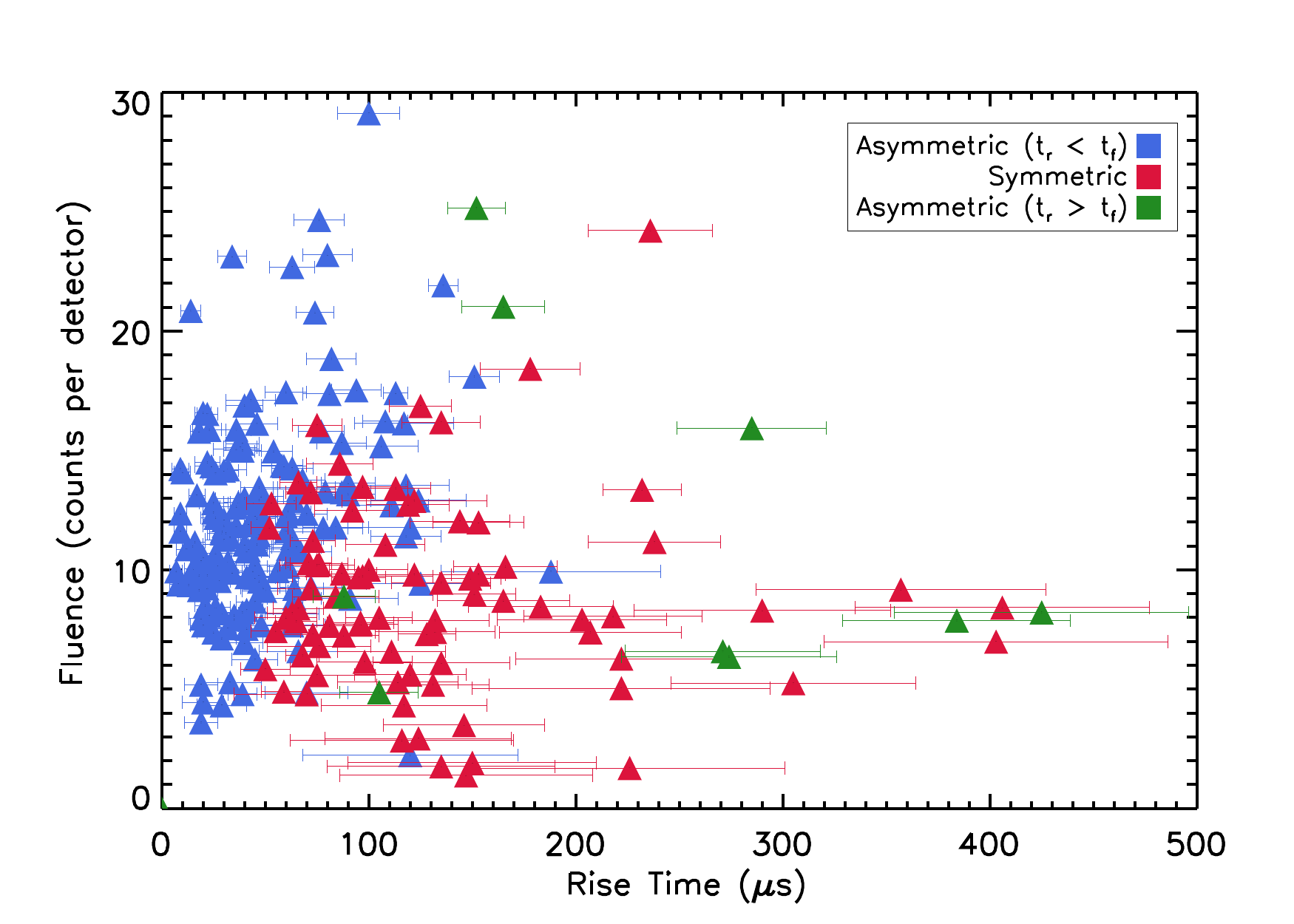}
\caption{Fluence versus rise time. The observed fluence is a lower limit on the actual value due to instrumental effects. There is a slight tendency for pulses with longer rise times to have lower fluences, consistent with the RFD model. However, pulses with shorter rise times have fluences which span the observed range.}
\label{fig:fluence_plots}
\end{figure}

\begin{figure} \centering
	\includegraphics[width=25pc]{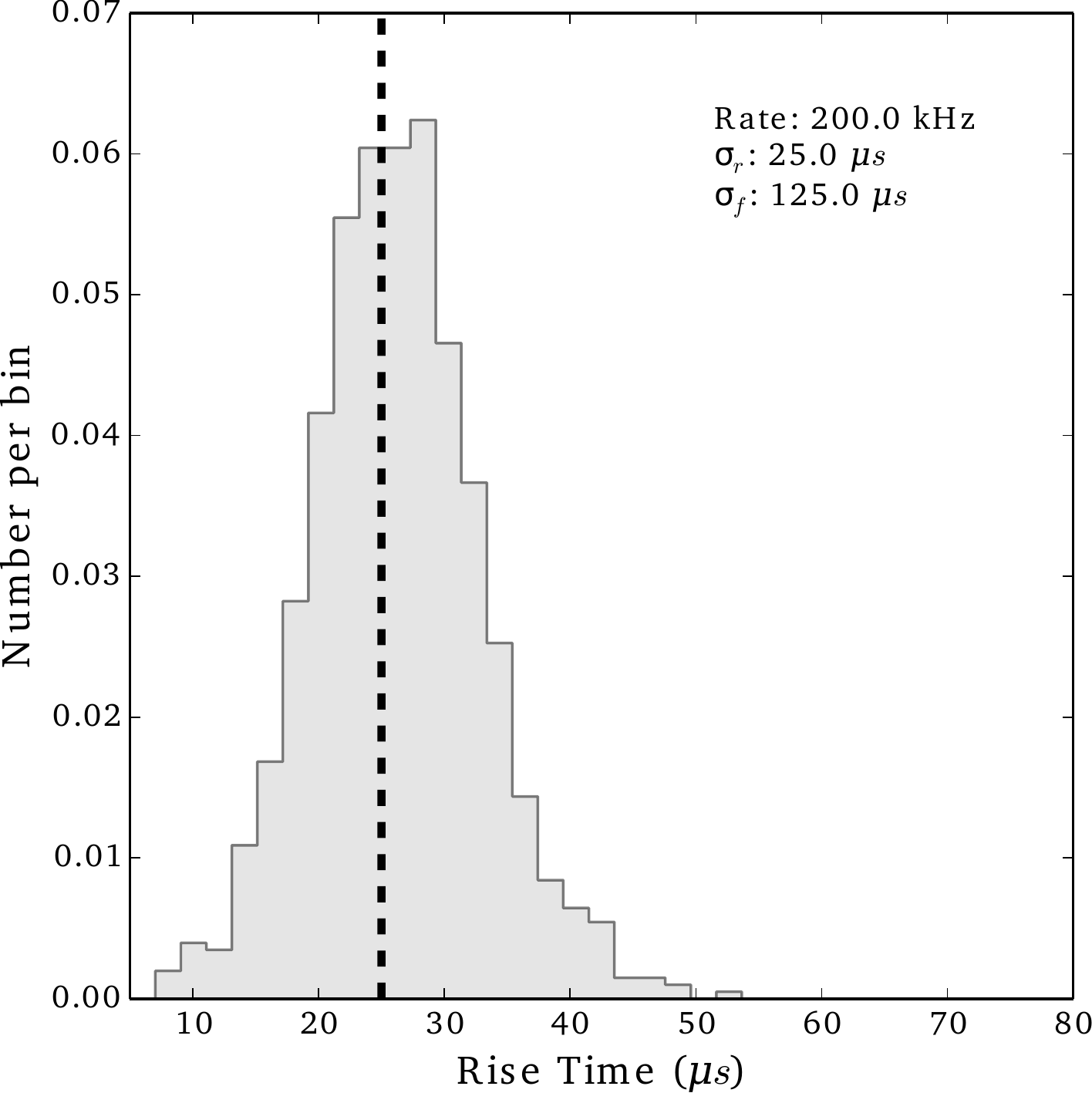}
\caption{Distribution of observed rise times for a strong (input peak rate of 200\,KHz per detector) asymmetric TGF. The input simulation rise time is plotted as vertical black line. The observed times are completely consistent with the input.}\label{fig:dt_rise}
\end{figure}

\begin{figure} \centering
	\includegraphics[width=25pc]{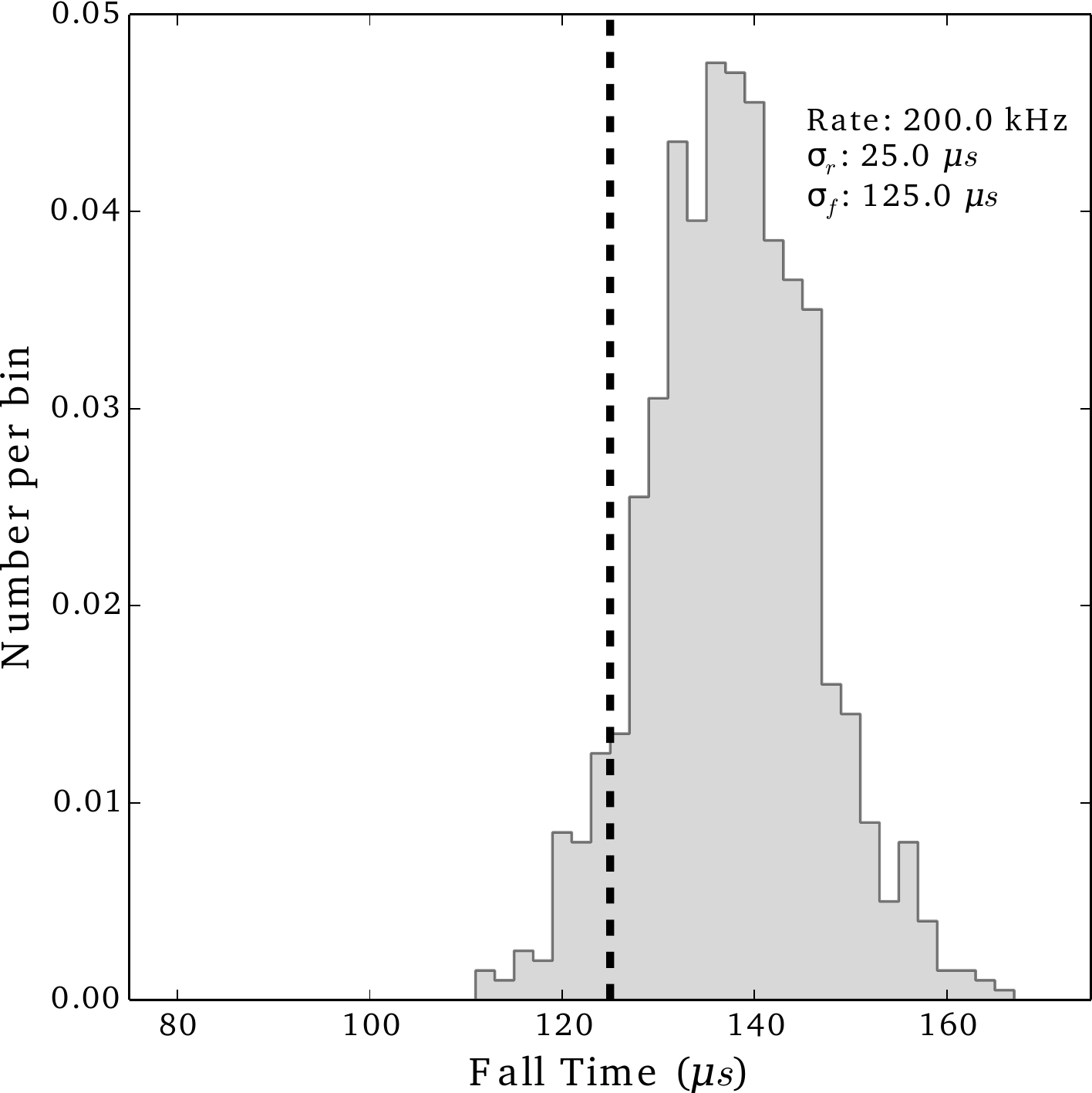}
\caption{Distribution of observed fall times for a strong (input peak rate of 200\,KHz per detector) asymmetric TGF. The input simulation fall time is plotted as vertical black line. The observed times are significantly shifted to longer times. However, the effect is relatively small at $\sim$10\,\%.}\label{fig:dt_fall}
\label{dt_fall}
\end{figure}

\begin{figure} \centering
	\includegraphics[width=25pc]{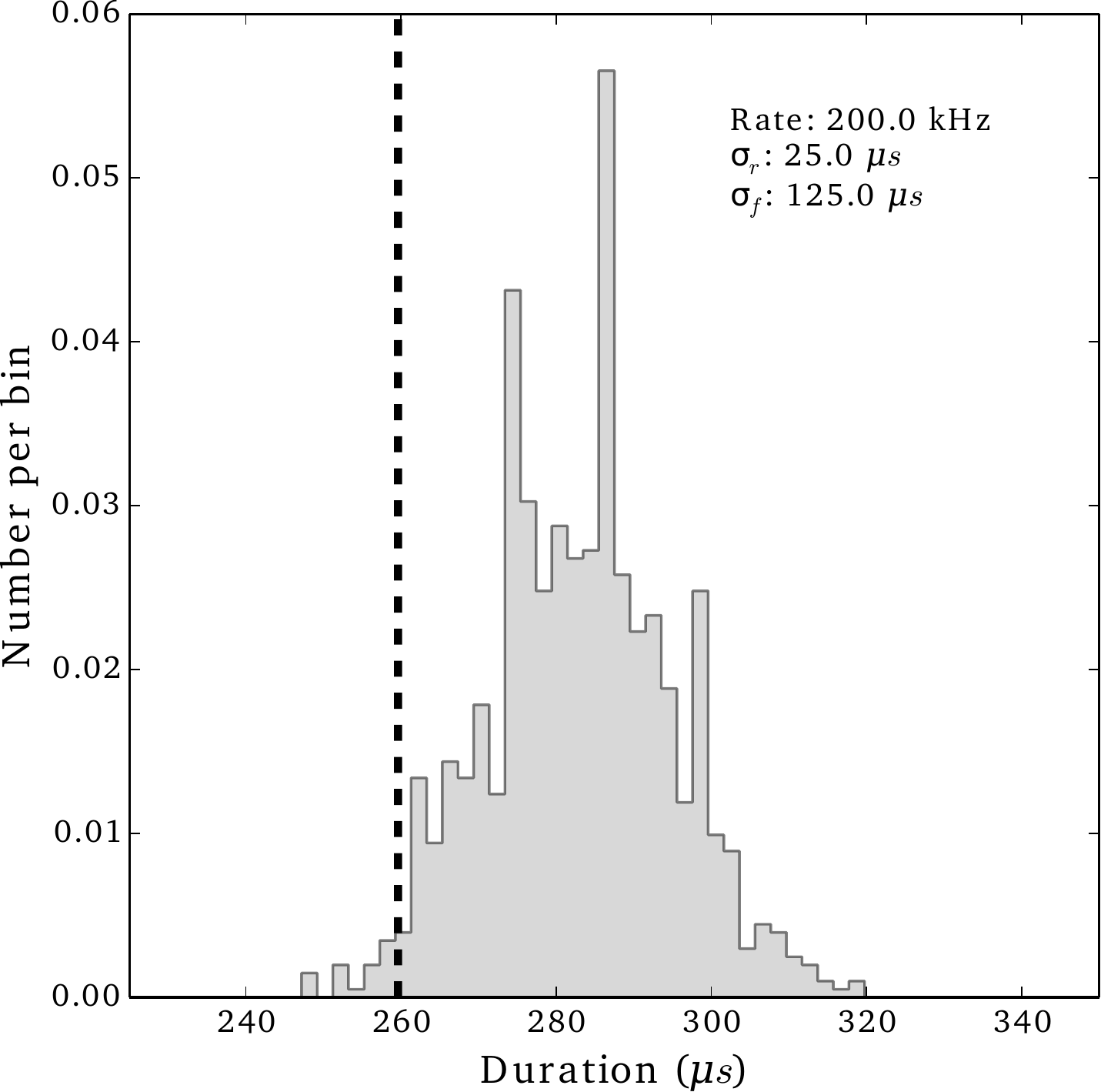}
\caption{Distribution of observed durations for a strong (input peak rate of 200\,KHz per detector) asymmetric TGF. The measured duration is clearly shifted to longer times, but the effect is small, with the mean being shifted by $\sim$\,20\,$\mu s$.}\label{fig:dt_duration}
\end{figure}

\begin{figure} \centering
	\includegraphics[width=40pc]{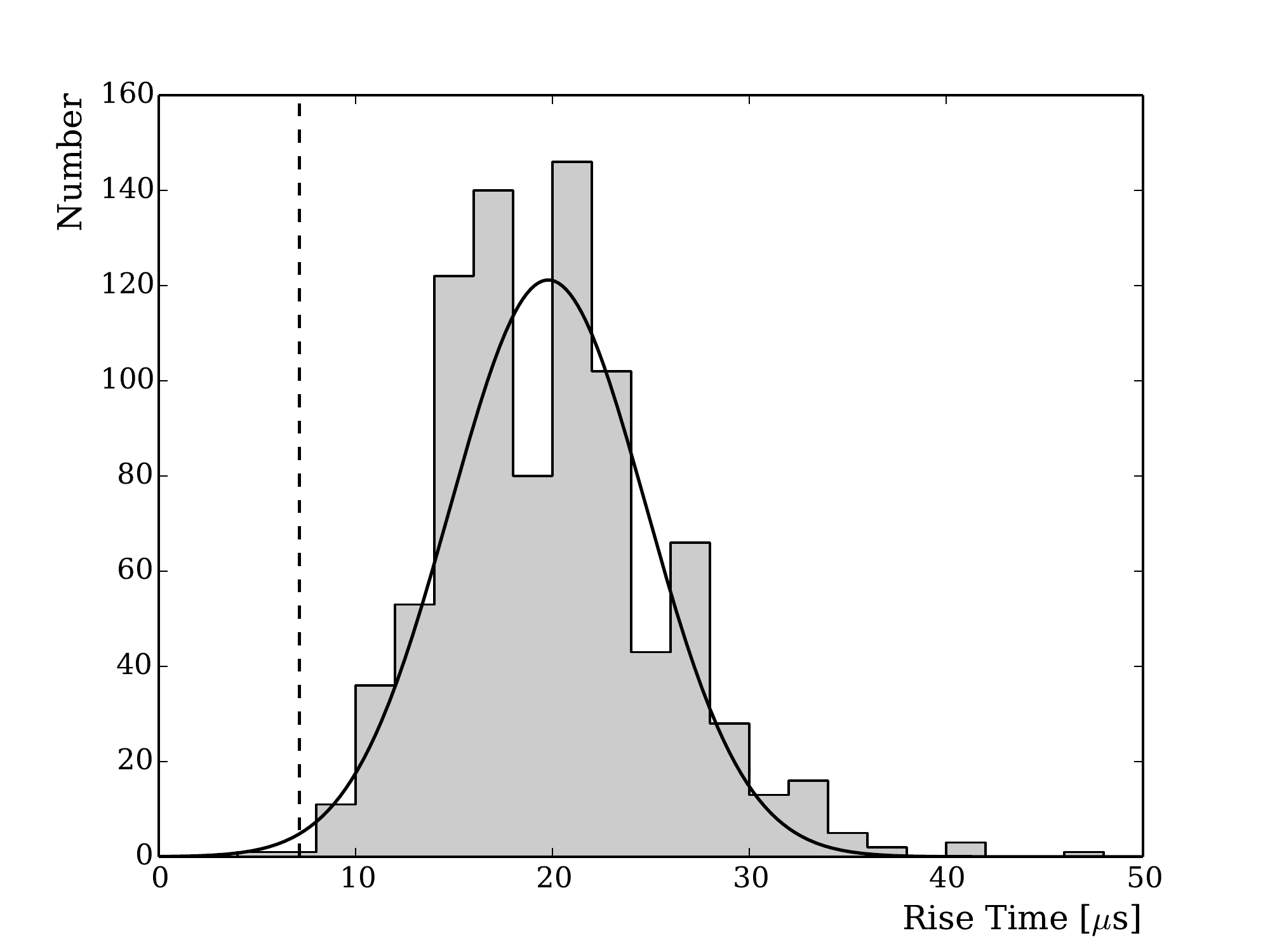}
\caption{Distribution of rise times for a strong TGF with a short rise time ($\sigma_r=15$\,$\mu$s). The solid line is the result of a Gaussian fit to the data and the dashed vertical line indicates the 7\,$\mu$s value. The probability of the 7\,$\mu$s value being drawn from the distribution is $\sim$0.5\%. }\label{fig:sim_rise}
\end{figure}

\begin{figure} \centering
	\includegraphics[width=40pc]{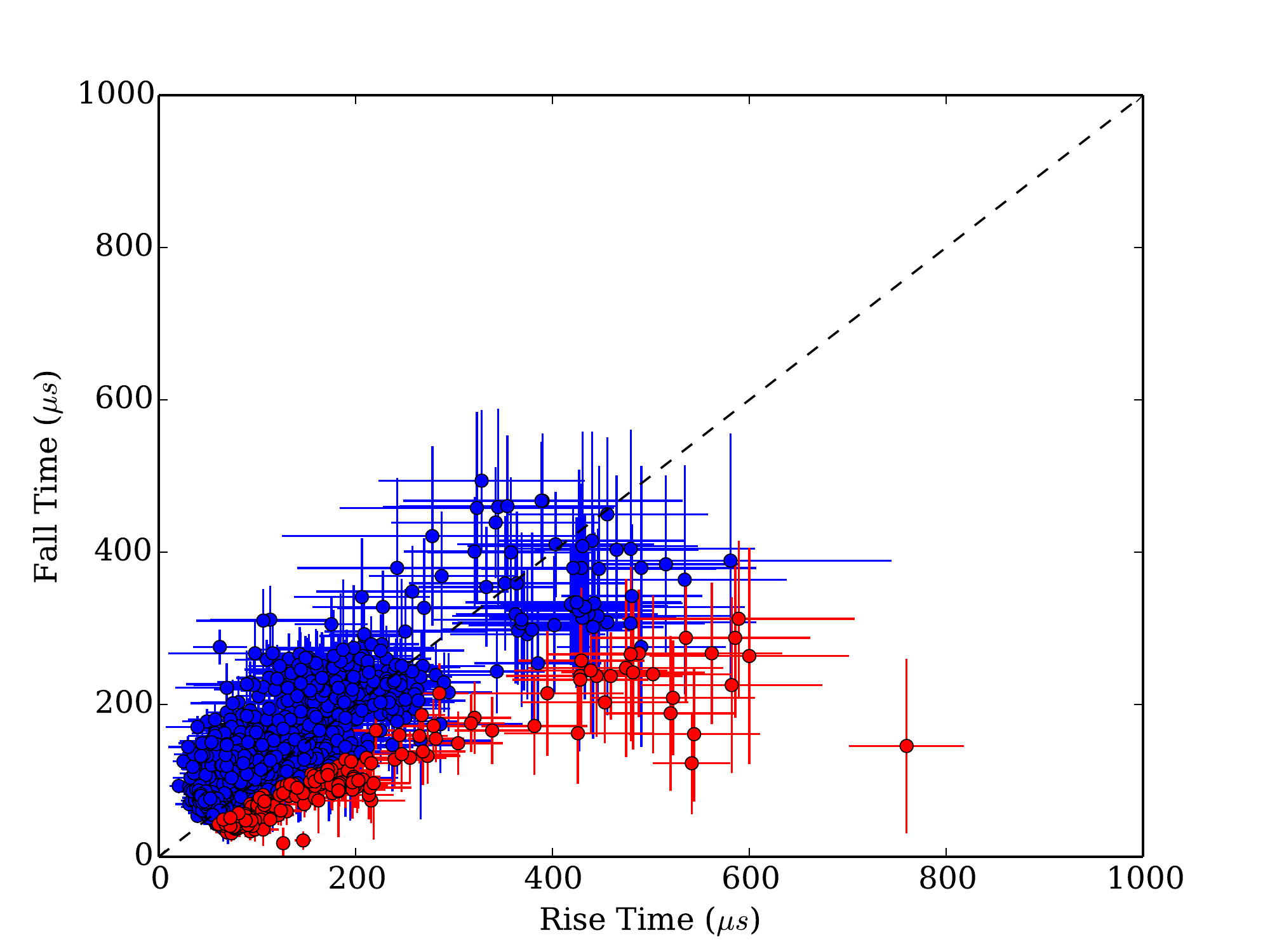}
\caption{Rise time vs fall time for 20 symmetric TGFs, each simulated 100 times. Points consistent with symmetry are shown in blue, those with longer rise times than fall times are shown in red. Approximately 12\,\% of the sample have longer rise time than fall time. }\label{fig:sim_riseFall}
\end{figure}

\end{document}